%% file: main.tex
\documentclass{article}
\usepackage[utf8]{inputenc}
\usepackage{amsmath, amssymb, amsthm}
\usepackage{mathrsfs}
\usepackage{cite}
\usepackage{hyperref}
\usepackage{authblk}
\usepackage{multirow}
\usepackage[table]{xcolor}
\usepackage{tikz}
\usepackage{algorithmic}
\usepackage[vlined, ruled, boxed]{algorithm2e}
\usepackage{adjustbox}
\usepackage{booktabs}
\usepackage{changepage}
\usepackage{qcircuit}
\usepackage{subcaption}

\newcommand{\pgroup}[0]{\mathcal{P}^*_n}
\newcommand{\pplus}[0]{\mathcal{P}^+_n}
\newcommand{\pauli}[0]{\mathcal{P}_n}

\newtheorem{definition}{Definition}

\SetKwProg{Fn}{Function}{}{}

\title{Faster and shorter synthesis\\ of Hamiltonian simulation circuits}
\author[1]{Timothée Goubault de Brugière}
\author[2]{Simon Martiel}
\affil[1]{Quandela, 7 Rue Léonard de Vinci, 91300 Massy, France}
\affil[2]{IBM Quantum, IBM France Lab, Orsay, France}

\begin{document}
\maketitle

\begin{abstract}
  We devise greedy heuristics tailored for synthesizing quantum circuits that implement a specified set of Pauli rotations. Our heuristics are designed to minimize either the count of entangling gates or the depth of entangling gates, and they can be adjusted to either maintain or loosen the ordering of rotations. We present benchmark results demonstrating a depth reduction of up to a factor of 4 compared to the current state-of-the-art heuristics for synthesizing Hamiltonian simulation circuits. We also show that these heuristics can be used to optimize generic quantum circuits by decomposing and resynthesizing them.
\end{abstract}

\section{Introduction}

Presently, quantum hardware faces challenges stemming from gate error rates and decoherence times. These technological constraints render the hardware unsuitable for quantum error correction but may still be sufficiently capable for applications in the so-called quantum utility regime, involving small quantum circuits with few qubits and low depth. Among the various algorithms and experiments developed for such devices, many are dedicated to quantum chemistry, condensed matter applications, or combinatorial optimization, including the famous VQE algorithm \cite{peruzzo2014variational}, the recent quantum utility experiments \cite{kim2023evidence}, or the seminal work on QAOA \cite{farhi2014quantum}. They involve implementing sequences of Pauli rotations with arbitrary angles, usually derived from first-order Trotter expansions of sparse Hamiltonians. It is common practice to overlook the non-commutativity of rotations in these sequences.

In this context, the goal is to determine the most cost-effective circuit implementation, typically the shallowest one, for a given set of Pauli rotations, with the flexibility to change the rotation order. The problem of minimizing depth or gate count is well-established in the literature, often focusing on subsets of circuits like linear Boolean reversible circuits (Maslov et al. \cite{maslov_zindorf}, de Brugière et al. \cite{debrugière2023shallower,tgdbdepth}) or phase polynomials (Amy et al. \cite{Amy_2018}). Several attempts have been made to develop effective heuristics for synthesizing sequences of Pauli rotations (van den Berg et al. \cite{van_den_Berg_2020}, Cowtan et al. \cite{cowtan2020generic}, Li et al. \cite{li2022paulihedral}), typically relying on local rewriting or other synthesis techniques.

In this work, we introduce two greedy heuristics to address this problem, targeting either the entangling gate count or the entangling gate depth of the circuit implementation. Unlike state-of-the-art methods that group Pauli rotations into pairwise commuting sets, codiagonalize them with a possibly large Clifford circuit, and then implement them with a phase polynomial algorithm (resulting in additional entangling gates), our approach simultaneously addresses these steps. We greedily select rotations that require the smallest Clifford circuits to be implemented trivially with a single qubit rotation. Some choices of Clifford circuits make subsequent Pauli rotations easier to implement as well. With well-defined cost functions, we identify efficient changes of basis and demonstrate that our heuristics outperform the state of the art for small instances, remaining competitive for larger instances while maintaining a low execution time.

Furthermore, we extend these heuristics to the order-preserving setting, enabling their use in generic circuit synthesis applications. Our algorithms remain competitive when compared to the state of the art, particularly for small instances where the greedy behavior captures optimizations. Overall, our algorithm complements methods tailored for large structures, such as codiagonalization \cite{de2022graph} and phase polynomial synthesis \cite{vandaele2022phase, Amy_2018}.

\paragraph{Outline} Section \ref{sec:prelim} is dedicated to some preliminaries and introduces the data structure used to represent Pauli rotations. Sections \ref{sec:low_count} and \ref{sec:low_depth} present our two main methods. Section \ref{sec:order_preserving} extends those methods to the order preserving setting. Finally section \ref{sec:benchmarks} presents some implementation details and benchmarks.

\section{Problem and framework}\label{sec:prelim}

We start with some preliminaries for the problem formalization.

\subsection{Preliminaries}

Pauli operators are specified by the four following Pauli matrices:

\

\begin{tabular}{cccc}
  $I=\begin{pmatrix}1&0\\0&1\end{pmatrix}$, & $X=\begin{pmatrix}0&1\\1&0\end{pmatrix}$, &$Y=\begin{pmatrix}0&-i\\i&0\end{pmatrix},$&$Z=\begin{pmatrix}1&0\\0&-1\end{pmatrix}.$ 
\end{tabular}

\

These matrices are pairwise anti-commuting involutions. Taking all possible tensor products of $n$ elements of the Pauli matrices times a $\pm 1/\pm i$ phase generate the full $n$-qubit Pauli group $\pauli$. We will focus on a particular subset of the Pauli group $\pplus \subset \mathcal{P}^*_n \subset \mathcal{P}_n$ containing operators with an overall phase $1$ and $\pm 1$ respectively. Given some operator $P\in\mathcal{P}_n$, we will denote $P(i)$ the $i$th Pauli matrix of $P$. Given some Pauli operator $P\in\pgroup$ and some angle $\theta \in \mathbb{R}$, we define the Pauli rotation $R_P(\theta)$ as the following unitary operator:
$$R_P(\theta) = \operatorname{exp}\left(-i\frac{\theta}{2}P\right) = \cos{(\theta/2)}I -i\sin{(\theta/2)}P.$$

The Clifford group $\mathcal{C}_n$ is defined as the set of unitary operators stabilizing $\mathcal{P}_n$:

$$\mathcal{C}_n = \{U\in SU(2^n) | U^\dagger P U \in \mathcal{P}_n, \forall P \in \mathcal{P}_n\}$$

By definition, Clifford operators weakly commute with Pauli rotations in the sense that for a given Clifford operator $U$ and Pauli rotation $R_P(\theta)$, there exists a Pauli operator $P'=U^{\dag}PU$ such that:
$$R_P(\theta)U =U R_{P'}(\theta).$$

Given some Pauli operator $P$, we call its \textit{support size} the number of its non-I terms. For instance, the $5$-qubit Pauli operator $I \otimes Z \otimes X \otimes I \otimes Z$ has support size $3$.

\paragraph{Vectorizing Pauli operators.} We now introduce the standard data structure used to represent Pauli operators and update them via conjugation by local Clifford gates. By considering the following $2-$bit encoding of Pauli generators
$$I = (0, 0)~~~~X = (0, 1)~~~~Z=(1, 0)~~~~Y=(1, 1)$$
we can represent any Pauli operator $P$ in $\pplus$ using a vector in $v\in\mathbb{F}_2^{2n}$, where the $(v[i], v[n+i])$ encodes the Pauli matrix $P(i)$. For instance, operator $XIYZ \in \mathcal{P}^+_4$ is encoded as :
$$ XIYZ = \begin{bmatrix}
    0 & 0& 1& 1&|&1& 0& 1& 0
\end{bmatrix}^T $$

We will call the first $n$ bits of the encoding the $Z$ part of the encoding, while the last $n$ bits will be called the $X$ part of the encoding. 

Given the encoding of some Pauli operator $P\in\pplus$, it is possible to compute in constant time the encoding of its image (up to a global phase in $\pm 1$) via conjugation by one or two qubit Clifford gate. In practice we will focus on gates $CNOT, H, R_X(\pi/2)=\sqrt{X}$, and $S$, whose actions on vectorized Pauli operators are described below. This gate set is (more than) universal for Clifford operators and suffices for our needs, our goal being to optimize CNOT count and/or depth.

\paragraph{Conjugation by a CNOT.}\label{paragraph:conj_cnot} Given some Pauli encoding $v\in \mathbb{F}_2^{2n}$ encoding some operator $P$, the encoding $v'$ of operator $CNOT_{i,j}\cdot P\cdot CNOT_{i,j}$ is such that:
\begin{itemize}
    \item $\forall k \notin\{j, n+i\}, v'[k]=v[k]$
    \item $v'[i] = v[j]\oplus v[i]$ and $v'[n+j] = v[n+i] \oplus v[n+j]$
\end{itemize}

\paragraph{Conjugation by a H.} Given some Pauli encoding $v\in \mathbb{F}_2^{2n}$ encoding some operator $P$, the encoding $v'$ of operator $H_i \cdot P\cdot H_i$ is such that:
\begin{itemize}
    \item $\forall k \notin\{i, n+i\}, v'[k]=v[k]$
    \item $v'[i] = v[n+i]$ and $v'[n+i] = v[i]$
\end{itemize}
In other words, conjugating with a H gate on qubit $i$ swaps the $i$th $X$ and $Z$ components of the encoding.

\paragraph{Conjugation by a $\sqrt{X}$.} Given some Pauli encoding $v\in \mathbb{F}_2^{2n}$ encoding some operator $P$, the encoding $v'$ of operator $\sqrt{X} \cdot P\cdot \sqrt{X}^\dagger $ is such that:
\begin{itemize}
    \item $\forall k \notin\{n+i\}, v'[k]=v[k]$
    \item $v'[n+i] = v[n+i] \oplus v[i]$
\end{itemize}
Thus, conjugating with a $\sqrt{X}$ gate on qubit $i$ sums the $i$th $Z$ component into the $i$th $X$ component.

\paragraph{Conjugation by a $S$.} Given some Pauli encoding $v\in \mathbb{F}_2^{2n}$ encoding some operator $P$, the encoding $v'$ of operator $S \cdot P\cdot S^\dagger $ is such that:
\begin{itemize}
    \item $\forall k \notin\{n+i\}, v'[k]=v[k]$
    \item $v'[i] = v[n+i] \oplus v[i]$
\end{itemize}
Thus, conjugating with a $S$ gate on qubit $i$ sums the $i$th $X$ component into the $i$th $Z$ component.

Notice that it is also possible to track the overall phase picked up during the conjugation (also in constant time). We don't showcase it here to not over complexify the framework. In the rest of this work, we will only consider Pauli operators in $\pplus$ and will ignore the potential $\pm 1$ phase that might appear.

\paragraph{Encoding a set of Pauli operators.} It is also useful to describe a set $S=\{P_1, ..., P_m\}$ Pauli operators in $\pplus$ as a table $M \in \mathbb{F}_2^{2n,m}$ where $M = \left[v_1 \cdots v_m\right]$ with $v_i$ the encoding of $P_i$.
In the same way, one can update in linear time this representation to amount for conjugation by a CNOT, H, $\sqrt{X}$, or $S$ gate: component additions (resp. swaps) becomes row additions (resp. swaps).
Even though this data structure does not change the asymptotic cost of conjugation, this allows for the vectorization of the conjugation process which greatly reduces the prefactor.

\subsection{Pauli network synthesis}

We are interested in implementing a set of Pauli rotations in any possible order. A natural way of formulating this problem is to generalize the notion of parity networks introduced in \cite{Amy_2018}. In the parity network setting, one tries to produce a CNOT+RZ circuit implementing a set of diagonal Pauli rotations. The problem is formulated as finding a CNOT circuit such that, for each rotation in our input set, there exists a time during the execution of the circuit where the rotation acts on a single qubit. Hence, the output CNOT circuit can be seen as a skeleton in which we will place some single qubit $Z$ rotations to implement the input set of rotations.

We propose here a generalization of a parity network, where the skeleton is not only composed of CNOT gates but of any Clifford gate.
Our goal is to produce a Clifford circuit, seen as a sequence of basis changes, such that each rotation becomes simple to implement at some point during its execution. In this setting, we will try to turn Pauli rotations into single qubit rotations, i.e. we will try to reduce the size of support of the rotations down to $1$.

\begin{definition}[Pauli network]
    A Clifford circuit $C$ is a Pauli network for a set $\{P_1, ..., P_m\}$ of $m$ Pauli operators if and only if there exists $m$ non-negative integers $p_{1}, \ldots, p_{m}$ such that $U_{i}^\dag P_i U_{i}$ has support size $1$, where $U_{i}$ is the Clifford operator implemented by the first $p_i$ gates of $C$.
\end{definition}

Notice that in this definition, we don't impose any particular ordering on the $p_i$ and thus on the order of appearance of the rotations in the output circuit. We can also formulate a similar definition to define networks which allow for the sequence to be implemented in particular order:
\begin{definition}[Ordered Pauli network]
    A Clifford circuit $C$ is a Pauli network for a sequence $(P_1, ..., P_m)$ of $m$ Pauli operators if and only if there exists $m$ non-negative integers $p_{1}, \ldots, p_{m}$ such that $U_{i}^\dag P_i U_{i}$ has support size $1$, where $U_{i}$ is the Clifford operator implemented by the first $p_i$ gates of $C$, and such that $\forall 1\leq i < j \leq m$, if $[P_i, P_j] \neq 0$, then $p_i \leq p_j$.
\end{definition}

Observe that the input is now an ordered sequence of operators, and that the $p_i$ are ordered in a way that preserves the structure of the sequence: only the order of commuting Pauli operators can be exchanged.


In this work, we are interested in producing low CNOT-count and low CNOT-depth (possibly ordered) Pauli networks for some some given set or sequence of Pauli operators. Solving this problem trivially provides us with a tool to synthesize circuits implementing a given (possible ordered) set of Pauli rotations. Moreover, coupled with rotation extraction and optimization techniques as introduced in \cite{zhang2019optimizing,martiel2022architecture}, such a synthesis algorithm can be used to optimize any input quantum circuit.

\paragraph{On the hardness of those problems.} When restricted to diagonal Pauli operators, those problems coincide and correspond to the \textit{parity network synthesis problem} introduced in \cite{Amy_2018}. In the same work, the authors show that this problem is hard in the \textit{encoded input case}, and conjectured it to be hard in the general case.

\section{Low entangling count synthesis}\label{sec:low_count}

In this section, we will attach ourselves to formulate a heuristic that attempts to produce (possibly ordered) Pauli network with a low CNOT-count. 

Our approach will rely on growing a Pauli network, starting from an empty circuit. At each step of the algorithm, we will enumerate a list of possible chunks of circuit to append to the current Pauli network. The choice of the chunk will be made using some score function.

Overall, the heuristic can be described by Algorithm \ref{alg:low_count}.

\begin{algorithm}[h!]
\caption{CountSynthesis}
\label{alg:low_count}
\begin{algorithmic}[1]
\REQUIRE A table $M=\left[P_1 \cdots P_m\right]$ of Pauli operator encodings
\ENSURE $C$ is a $\{CNOT, \sqrt{X}, H\}$-Pauli network for $\{P_1, ..., P_m\}$
\STATE $C\gets [\ ]$
\WHILE{$M$ is not empty}
\STATE SortColumns($M$)
\WHILE{$M[:, 0]$ has support size $1$}
\STATE pop the first column of $M$
\ENDWHILE
\STATE $S \gets \operatorname{Support}(M[:, 0])$
\STATE $D_{\operatorname{max}} = \operatorname{argmax}_{D \in \mathcal{D}, i < j\in S} Score(M, D, i, j)$
\STATE $M\gets D_{\operatorname{max}}^{(i,j)}\cdot M \cdot {D_{\operatorname{max}}^{(i,j)}}^\dagger$
\STATE C.extend$(D_{\operatorname{max}})$
\ENDWHILE
\end{algorithmic}
\end{algorithm}

In other words, the heuristic proceeds as follows:
\begin{itemize}
    \item the columns of $M$ are sorted in ascending support size (l.3)
    \item the ``trivial'' columns of $M$ are removed (l.4-6): they correspond to rotations that are already synthesized by the current Pauli network $C$
    \item a collection of small CNOT+H+$\sqrt{X}$ circuits (or \textit{chunks}) $D\in \mathcal{D}$ are enumerated for each ordered pair of qubits $i\neq j\in S$ where $S$ is the support of the leading rotation.
    \item a score is computed for each choice of $D,i,j$. The choice maximizing the score function is retained and used to further extend the circuit $C$. $M$ is updated accordingly(l.8-10).
\end{itemize}

We now make explicit the two ingredients of this heuristic that are the score function and the collection of chunks $\mathcal{D}$.

 \paragraph{Score function.} First, let us explicit the score function used at line 8. This function has access to the table $M$, the chunk $D$ and a pair of qubits $i\neq j$. 
 This score function relies on computing the number of leading $0$s appearing on the modified rows of $M$ (i.e. rows $\{i, n+i, j, n+j\}$). For this purpose, we define the following helper function: 
 $$\operatorname{CLI}(M, q) = \operatorname{min}(\operatorname{leading\_0s}(M[q,:]), \operatorname{leading\_0s}(M[n+q,:]))$$
 Then, our score function will simply compute the difference $$\operatorname{CLI}(D^{(ij)}\cdot M\cdot {D^{(ij)}}^{\dagger}, q) - \operatorname{CLI}(M, q)$$ for $q \in \{i, j\}$.
 In other words, it counts how many leading identities were added on qubits $i$ and $j$ when conjugating the set of rotations by our candidate chunk.
 
Notice that since identities can only appear if some operator $P_i\otimes Q_j$ was turned into $P'_i\otimes I$ or $I\otimes Q'_j$, this number of leading identities can only vary for either $i$ or $j$ but not both. For this reason, we could equivalently track the sum of the scores of the two qubits, or the maximum between the two scores, without changing the behavior of our score function.

In practice, we use the following score function:
$$\operatorname{Score}(M,D,i,j)=\operatorname{max}_{q\in\{i,j\}}\left(\operatorname{CLI}(D^{(ij)}\cdot M\cdot {D^{(ij)}}^{\dagger}, q) - \operatorname{CLI}(M, q)\right)$$

\paragraph{Clifford chunks.} Let us now introduce the family $\mathcal{D}$ of chunks enumerated at l.8.
Our goal is to find a family of single CNOT piece of Clifford circuit to add to our Pauli network that can increase the value of our score function. Since our function computes the number of \emph{added leading identities}, we want our chunks to be able to increase this quantity whenever it is possible. In particular, for any pair of two-qubit Pauli operators $A\otimes B$ and $P\otimes Q$, if there exists a single-CNOT Clifford circuit that transforms them into the form $A'\otimes I$ and $P'\otimes I$, our objective is to identify this circuit.

Let us assume that $A\otimes B=Z\otimes Z$. Then $P\otimes Q$ is co-reducible to a $I\otimes Q'$ via a single CNOT gate if and only if the following condition holds: 
\begin{align}
    P=Z\wedge Q \in \{Y,Z\}\label{eq:cond_cored_cnot}
\end{align}
This fact is easy to check using the CNOT conjugation rule (c.f. Paragraph \ref{paragraph:conj_cnot}). Moreover, since the only non-trivial single-qubit Clifford gate that commutes with $Z$ is $S$, one can also co-reduce pairs $P'\otimes Q'$ such that:
\begin{align}
    P'=Z\wedge Q \in \{S^{\dag b}YS^b, Z\}\label{eq:cond_cored_cnot_s}
\end{align}
for some $b\in\{0, 1\}$. Consequently, the only pairs that can be co-reduced with $Z\otimes Z$ using a circuit containing a single CNOT are of shape $Z\otimes \{X,Y,Z\}$. The cases $Z\otimes \{Y,Z\}$ are trivially handled by a CNOT gate (condition \ref{eq:cond_cored_cnot}), while the case $Z\otimes X$ is handled by $CNOT\cdot (I\otimes S)$.

For this reason, we consider chunks of shape:

\[
\begin{array}{c}
\Qcircuit @C=1em @R=.7em {
&\qw&\gate{U_1}&\qw&\ctrl{1}&\qw\\
&\qw&\gate{U_2}&\gate{S^b}&\targ&\qw\\
}
\end{array}
\]
where $U_i \in \{I, H, \sqrt{X}\}$. Thus, we pick: $$\mathcal{D}=\{CNOT\cdot (I\otimes S^b) \cdot (U_1 \otimes U_2), U_i \in \{I, H, \sqrt{X}\}, b\in \{0, 1\}\}$$

\begin{figure}
    \begin{subfigure}[t]{\linewidth}

\[ \Qcircuit @C=1em @R=.7em {
&\qw&\gate{S}&\ctrl{1}&\qw& \\ 
&\qw&\qw&\targ&\qw&\\
} = 
\Qcircuit @C=1em @R=.7em {
&&\qw&\ctrl{1}&\gate{S}&\qw \\ 
&&\qw&\targ&\qw&\qw\\
}
\]
\caption{Removing $S$ on control qubit.}
\end{subfigure}
\ \\

\begin{subfigure}[t]{\linewidth}
\[ \Qcircuit @C=1em @R=.7em {
&\qw&\gate{S}&\gate{H}&\ctrl{1}&\qw& \\ 
&\qw&\qw&\qw&\targ&\qw&\\
} = 
\Qcircuit @C=1em @R=.7em {
&&\qw&\gate{S}&\gate{H}&\gate{S}&\gate{S}&\ctrl{1}&\qw& \\ 
&&\qw&\qw&\qw&\qw&\qw&\targ&\qw&\\
} = 
\Qcircuit @C=1em @R=.7em {
&&\qw&\gate{\sqrt{X}}&\ctrl{1}&\gate{S}&\qw \\ 
&&\qw&\qw&\targ&\qw&\qw\\
}
\]
\caption{Removing $HS$ on control qubit.}
\end{subfigure}
\ \\

\begin{subfigure}[t]{\linewidth}
\[ \Qcircuit @C=1em @R=.7em {
&\qw&\gate{H}&\gate{S}&\ctrl{1}&\qw& \\ 
&\qw&\qw&\qw&\targ&\qw&\\
} = 
\Qcircuit @C=1em @R=.7em {
&&\qw& \gate{H}&\ctrl{1}&\gate{S}&\qw \\ 
&&\qw&\qw&\targ&\qw&\qw\\
}
\]   
\caption{Removing $SH$ on control qubit.}
\end{subfigure}
\
\\

\begin{subfigure}[t]{\linewidth}
\[ \Qcircuit @C=1em @R=.7em {
&\qw&\qw&\ctrl{1}&\qw& \\ 
&\qw&\gate{\sqrt{X}}&\targ&\qw&\\
} = 
\Qcircuit @C=1em @R=.7em {
&& \qw&\ctrl{1}&\qw&\qw\\ 
&& \qw&\targ&\gate{\sqrt{X}}&\qw&\\
}
\] 
\caption{Removing $\sqrt{X}$ on target qubit.}
\end{subfigure}
\
\\
  
\begin{subfigure}[t]{\linewidth}
\[ \Qcircuit @C=1em @R=.7em {
&\qw&\qw&\qw&\ctrl{1}&\qw& \\ 
&\qw&\gate{S}&\gate{H}&\targ&\qw&\\
} = 
\Qcircuit @C=1em @R=.7em {
&&\qw&\qw&\qw&\qw&\qw&\ctrl{1}&\qw& \\ 
&&\qw&\gate{H} & \gate{H} & \gate{S}&\gate{H}&\targ&\qw&\\
}
= 
\Qcircuit @C=1em @R=.7em {
&&\qw&\qw&\ctrl{1}&\qw&\qw \\ 
&&\qw&\gate{H}&\targ&\gate{\sqrt{X}}&\qw\\
}
\]  
\caption{Removing $HS$ on target qubit.}
\end{subfigure}
\ \\

\begin{subfigure}[t]{\linewidth}
\[ \Qcircuit @C=1em @R=.7em {
&\qw&\qw&\qw&\ctrl{1}&\qw& \\ 
&\qw&\gate{H}&\gate{S}&\targ&\qw&\\
} = 
\Qcircuit @C=1em @R=.7em {
&&\qw&\qw&\qw&\qw&\qw&\ctrl{1}&\qw& \\ 
&&\qw&\gate{S} & \gate{S} & \gate{H}&\gate{S}&\targ&\qw&\\
}
= 
\Qcircuit @C=1em @R=.7em {
&&\qw&\qw&\ctrl{1}&\qw&\qw \\ 
&&\qw&\gate{S}&\targ&\gate{\sqrt{X}}&\qw\\
}
\]  
\caption{Removing $SH$ on target qubit.}
\end{subfigure}

\caption{Commutation rules of one-qubit Clifford gates with a CNOT gate.}
\label{commutation}
\end{figure}
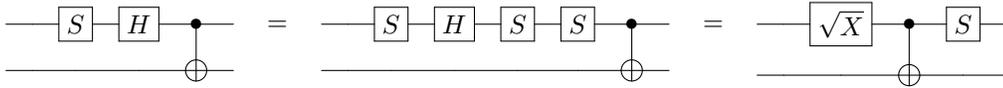
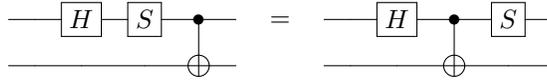
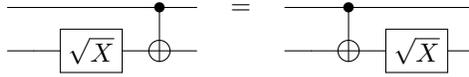
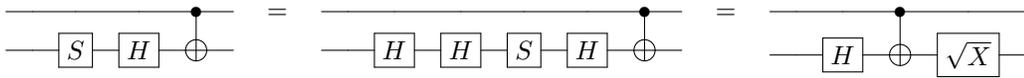
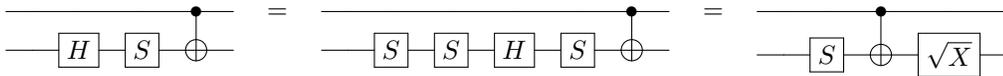

Ignoring phases, there are $6$ different 1-qubit Clifford circuits: $I$, $H$, $S$, $HS$, $SH$ and $\sqrt{X} = HSH = SHS$. Using commutation rules illustrated in Fig.~\ref{commutation}, we show that we can post-process some one-qubit Clifford which eventually do not modify the score function. Therefore some chunks are equivalent such that we only need
$$\mathcal{D}=\{CNOT\cdot (U_1\otimes U_2), \; \; U_1 \in \{I, H, \sqrt{X}\}, U_2 \in \{I, H, S\}\}$$
for a total of $2 \cdot 3^2 = 18$ possible chunks.

Line 8 of Algorithm \ref{alg:low_count} thus consists in enumerating all circuits in 
 this set $\mathcal{D}$ and all ordered pairs of distinct qubits that appear in the first rotation (there are ${w(w-1)}$ such pairs, where $w$ is the support size of the first rotation).


\paragraph{Complexity.} Assuming an instance over $n$ qubits with $m$ rotations to synthesize. 
The main loop of the algorithm is executed as many times as there are CNOT gates in the output circuit. This number can be upper bounded trivially by $O(nm)$. Inside the loop, the dominating step is l.8 where we compute the score of each pair of qubits. There are up to $O(n^2)$ pairs to consider and each score computation requires $O(m)$ operations. Thus the overall complexity is a
$O(n^3m^2)$.

\section{Low depth synthesis}\label{sec:low_depth}

 In this section, we describe another heuristic focused on trying to minimize the \textit{entangling depth} of the produced Pauli network.
The general approach will be roughly the same as the one of Algorithm \ref{alg:low_count} except that the heuristic will attempt to pick several chunks to apply on non intersecting pairs of qubits. In order to pick the best set of non intersecting chunks, the heuristic relies on the well known problem of maximal matching. In practice it will, at each iteration:
\begin{itemize}
    \item for each (unordered) pair of qubit $\{i, j\}$ compute the best chunk to apply and its direction (i.e. on $(i,j)$ or $(j,i)$)
    \item using this information, build a complete weighted graph
    \item use a standard blossom algorithm to compute a maximal weight matching of the resulting graph
    
\end{itemize}
The obtained matching will describe a depth-1 circuit that can be used to extend the current Pauli network. The pseudo-code of the algorithm is detailed in Algorithm \ref{alg:low_depth}.

\begin{algorithm}[h!]
\caption{DepthSynthesis}
\label{alg:low_depth}
\begin{algorithmic}[1]
\REQUIRE A table $M=\left[P_1 \cdots P_m\right]$ of Pauli operators encodings
\ENSURE $C$ is a $\{CNOT, \sqrt{X}, H\}$-Pauli network for $\{P_1, ..., P_m\}$
\STATE $C\gets [\ ]$
\WHILE{$M$ is not empty}
\STATE SortColumns($M$)
\WHILE{$M[:, 0]$ has support size $1$}
\STATE pop the first column of $M$
\ENDWHILE

\tcp{$A$ is the adjacency matrix of a weighted complete graph}
\STATE $A \gets $ a all-zero $n\times n$ matrix 
\FORALL{$i<j \in [1,n]$}
    \STATE $A[i,j] = \operatorname{max}_{D \in \mathcal{D}} Score(M, D, i, j)$
    \STATE $A[j,i]=A[i,j]$
\ENDFOR

\tcp{A call to blossom algorithm}
\STATE $L = SolveMatching(A)$
\FOR{$(D,i,j) \in L$}
\STATE $M\gets D^{(i,j)}\cdot M \cdot {D^{(i,j)}}^\dagger$
\STATE C.extend$(D)$
\ENDFOR
\ENDWHILE
\end{algorithmic}
\end{algorithm}

\paragraph{Complexity.} Building the complete weighted graph costs $O(n^2m)$. The maximum matching can be computed in $O(n^3)$ using Gabow's algorithm \cite{gabowmatching}. Overall, the time complexity of the full algorithm can be bounded by $O(n^3m^2+n^4m)$. 

\section{Extension to ordered Pauli network synthesis}\label{sec:order_preserving}

The two heuristics presented in Sections \ref{sec:low_count} and \ref{sec:low_depth} can be easily adapted for them to produce ordered Pauli networks.
For this purpose, we rely on a data structure introduced in \cite{zhang2019optimizing} that captures the anti-commuting relations in a sequence of Pauli operators. The structure consists of a Directed Acyclic Graph (DAG) $D=(V, E)$ such that $V$ is exactly the set of Pauli operators in our sequence and $(P_i, P_j)\in E$ if and only if $[P_i, P_j]\neq 0$ and $i < j$.
This data structure can be constructed in time $O(nm^2)$.

Once equipped with such a structure $D$, one can easily extract the set of vertices in $V$ that have no incoming edges. We call this set the \textit{front layer} of $D$. In this subset, the operators all pairwise commute and can be synthesized in any order.

Our strategy is summed up by Algorithm \ref{alg:generic_order_preserving}. In this pseudo-code, the \emph{SingleStep} refers to lines 3 to 10 of Algorithm \ref{alg:low_count} or lines 3 to 16 of Algorithm \ref{alg:low_depth}.

\begin{algorithm}[h!]
\caption{Ordered Pauli network synthesis}
\label{alg:generic_order_preserving}
\begin{algorithmic}[1]
\REQUIRE A table $M=\left[P_1 \cdots P_m\right]$ of Pauli operators encodings
\ENSURE $C$ is an ordered $\{CNOT, \sqrt{X}, H\}$-Pauli network for $\{P_1, ..., P_m\}$
\STATE $D\gets \operatorname{DAG}(M)$
\STATE $F \gets \operatorname{FrontLayer}(D)$ \tcp{Gets the front layer as a table of Pauli operators}
\STATE $C\gets [\ ]$
\WHILE{$F$ is not empty}
    \STATE C.extend$(SingleStep(F))$
    \STATE $D\gets \operatorname{Update}(D, F)$
    \STATE $F \gets \operatorname{FrontLayer}(D)$
\ENDWHILE
\end{algorithmic}
\end{algorithm}

\paragraph{Complexity.} The worst time complexity remains the same since the construction of the Pauli DAG is negligible.

\section{Implementation and benchmarks}\label{sec:benchmarks}

We implemented Algorithms \ref{alg:low_count} and \ref{alg:low_depth} and their order preserving extensions in Rust, binded in python for simpler interfacing with other packages. The code can be found on github\footnote{\href{https://github.com/smartiel/rustiq}{https://github.com/smartiel/rustiq}} (together with other synthesis algorithms for other classes of operators). In can be installed directly via pip and supports very simple input and output formats. Our implementation supports the synthesis of the final Clifford operator if required.

We compared our implementations (\emph{rcount} and \emph{rdepth}) with the pytket library  (\emph{tket}), the Paulihedral algorithm based on qiskit (\emph{ph}), and a naive synthesis where each rotation is synthesized individually by a naive CNOT ladder (\emph{naive}).

We ran those algorithms on two distinct sets of benchmarks which are detailed below.

\paragraph{UCCSD Ansätze} The first set of benchmark consists of all the inputs used in  \cite{cowtan2020generic}. Each input is a collection of Pauli operators corresponding to the rotation axis required for the implementation of a UCCSD Ansatz for a given molecule. 

\begin{figure}
    \centering
\begin{adjustwidth}{-50pt}{0pt}
    \includegraphics[scale=0.5]{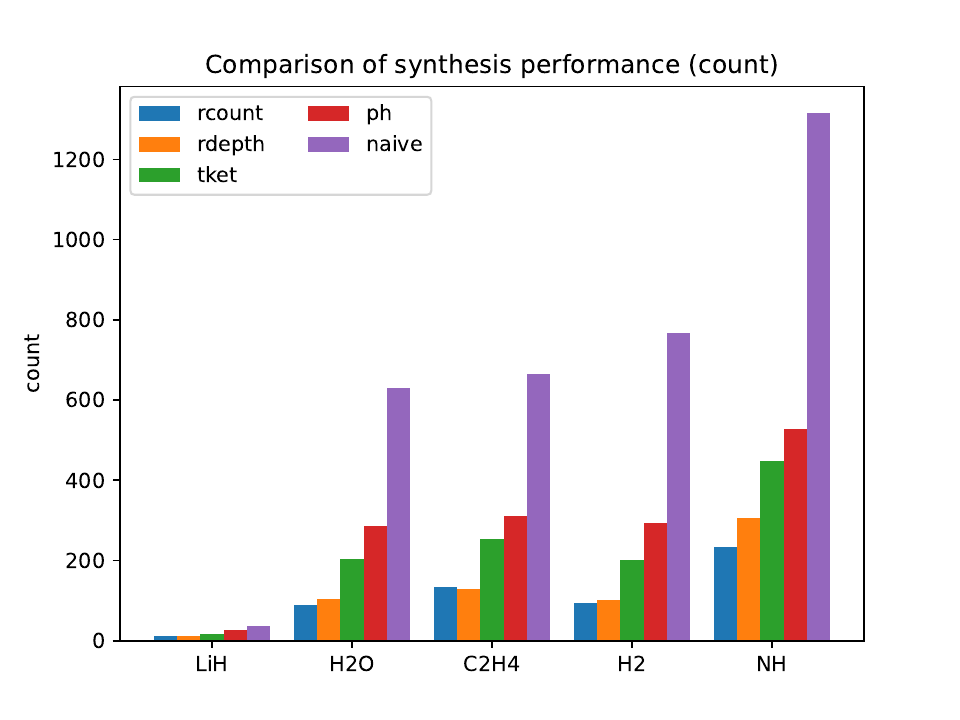}\includegraphics[scale=0.5]{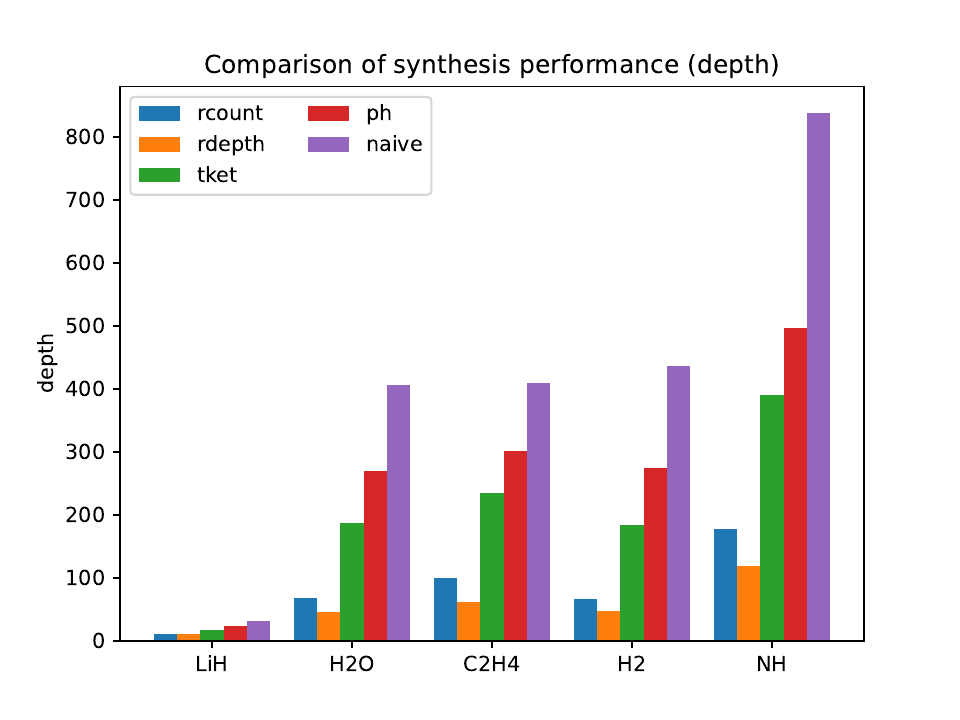}
\end{adjustwidth}
    
    \caption{Visual comparison of the CNOT count and depth of typical circuits produced by the various methods on small instances. Those instances correspond respectively to entries {\bf LiH cmplt JW sto3g},  {\bf H2O cmplt P sto3g},  {\bf C2H4 cmplt JW sto3g},  {\bf H2 cmplt P ccpvdz}, and  {\bf NH frz P sto3g} of Tables \ref{tab:tab_chem_1} to \ref{tab:tab_chem_depth_3}.}
    \label{fig:small_mol}
\end{figure}

Tables \ref{tab:tab_chem_1} to \ref{tab:tab_chem_depth_3} summarize the first set of benchmarks. Overall, both the depth and count heuristics are competitive at tackling small instances. When moving to larger instances, the count heuristic seems to be outperformed by the Paulihedral heuristic, while still improving upon tket in most cases. In the entangling depth setting, our depth heuristic strictly outperforms both state of the art methods by a healthy margin. Notice that our heuristic sometimes manages to reduce the output circuit depth by a factor up to 4 w.r.t. the \emph{tket} heuristic which was first introduced in \cite{cowtan2020generic} as a depth reduction heuristic (see row {\bf LiH frz JW ccpvdz} of Table \ref{tab:tab_chem_depth_1}). When compared to \emph{ph}, our depth heuristics still manages to reduce the circuit depth by a factor of up to 5.7 (entry {\bf H2 cmplt P ccpvdz} of Table \ref{tab:tab_chem_depth_1}). Figure \ref{fig:small_mol} gives a qualitative overview of the different synthesis outputs for small instances.

\input{table_new_4.tex}

\begin{figure}
    \centering
    \begin{adjustwidth}{-50pt}{0pt}
    \includegraphics[scale=0.5]{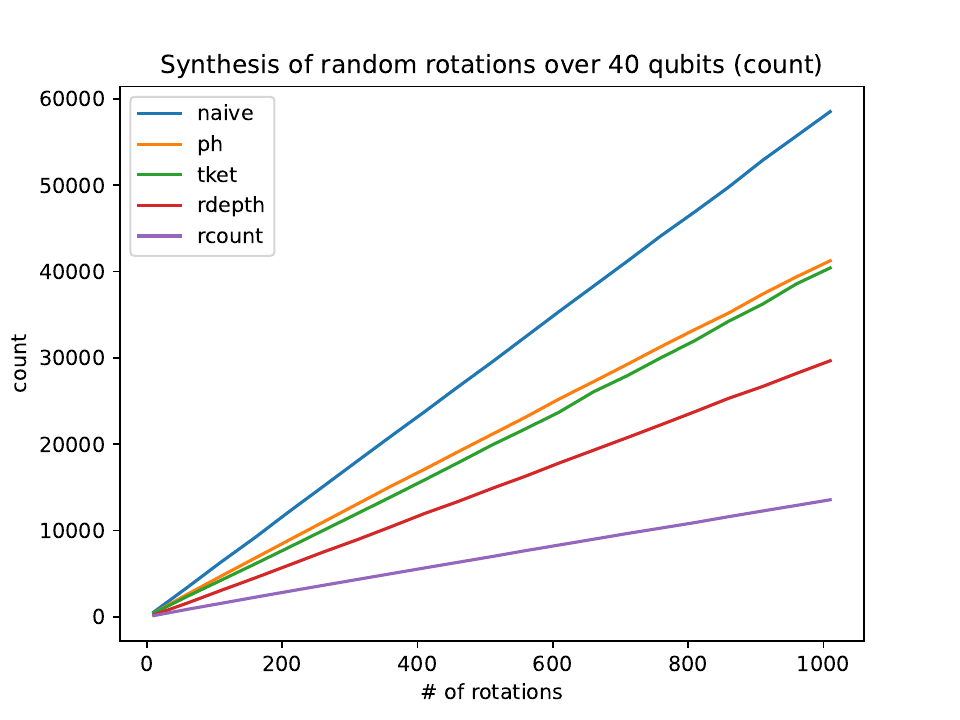}\includegraphics[scale=0.5]{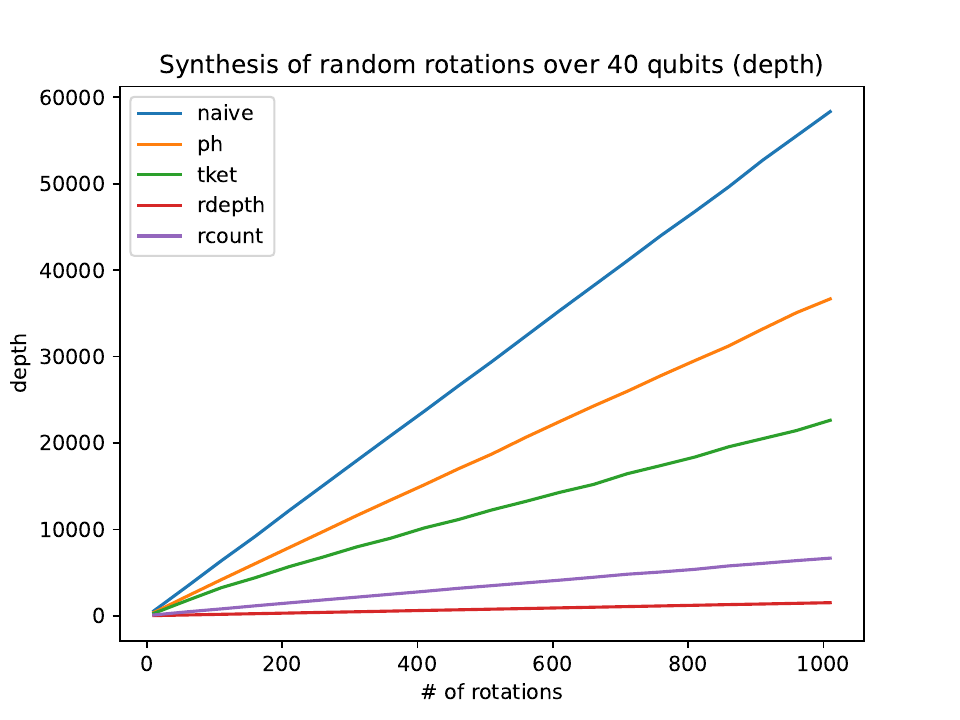}
    \end{adjustwidth}
    \caption{Average count and depth resulting from the synthesis of a sequence of rotations over 40 qubits. Each point is averaged over 10 instances.}
    \label{fig:rnd_bench}
\end{figure}

\begin{figure}
    \centering
    \begin{adjustwidth}{-50pt}{0pt}
    \includegraphics[scale=0.5]{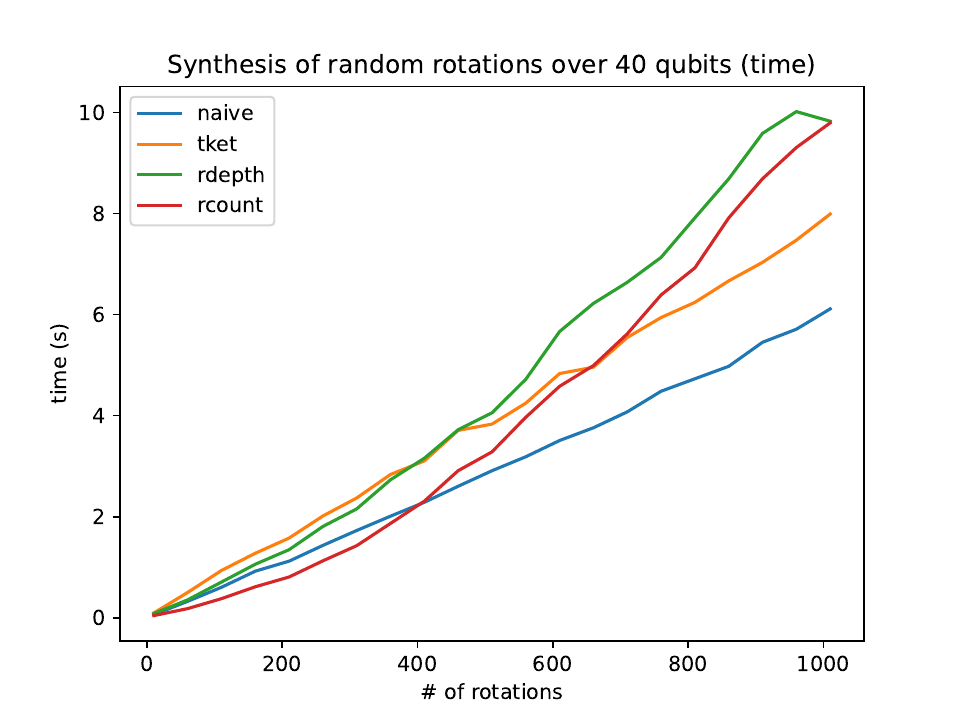}\includegraphics[scale=0.5]{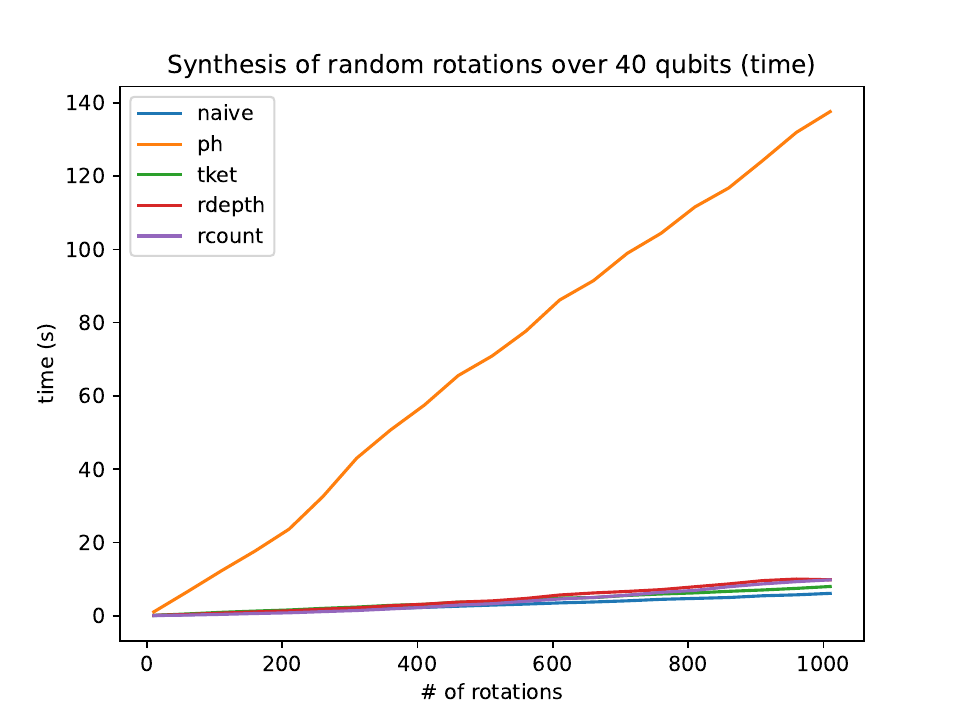}
    \end{adjustwidth}
    \caption{Average running time of the synthesis of a sequence of rotations over 40 qubits. Each point is averaged over 10 instances. Left data excluding method \emph{ph} for clarity. Right same data including \emph{ph}.}
    \label{fig:rnd_time}
\end{figure}

\paragraph{Random instances} The second set consists of random sequences of Pauli rotations. Even though this set is less practically relevant, it still provides a good idea of the scaling of the different heuristics. Figure \ref{fig:rnd_bench} presents those results. Notice that our ``depth'' heuristic outperforms all other heuristics both in entangling depth and time (see \ref{fig:rnd_time}).

\paragraph{Resynthesis of literature circuits.} Many CNOT or $T$ gate optimization algorithms were benchmarked against the same set of literature circuit. We decided to try and re-synthesize those circuits via our methods. Our protocol is the following:
\begin{itemize}
    \item express the input circuit $U$ as: $$U = C\cdot \prod_{i=1}^k R_{P_i}(\theta_i)$$ where $C$ is a Clifford operator and $R_{P_i}(\theta_i)$ are Pauli rotations,
    \item simplify the sequence $\prod_{i=1}^k R_{P_i}(\theta_i)$ using the T-count optimization algorithm from \cite{zhang2019optimizing},
    \item synthesize the resulting sequence of rotations via one of our heuristic in the order preserving setting (i.e. preserving the underlying unitary map).
\end{itemize}
The results can be found in Table \ref{tab:lit_circuits}. Our heuristic approach manages to outperform algorithms from \cite{Amy_2018} and \cite{Nam_2018} on 19 out of 39 benchmarks, with a tie on one occasion. This is noteworthy, as our method relies solely on greedy techniques, whereas the latter two algorithms employ global optimization of phase polynomials. Due to the lack of entangling depth metrics provided by \cite{Amy_2018, Nam_2018}, it is challenging to compare the performance of our optimized circuits in this regard. Observe that the circuits generated by the \emph{rdepth} method are shallower than the original circuits in all but 4 cases. Our findings suggest that our approach may be competitive for generic quantum circuit optimization, and further investigation is necessary to fully understand its applicability and potential advantages.

\input{table_resynth}

\section{Conclusion}

We introduce two heuristics designed to efficiently synthesize sequences of Pauli rotations with minimal entangling gate count or depth. Additionally, we offer variants of these heuristics that maintain rotation ordering through legal commutations.

Benchmarking demonstrates significant depth improvements compared to state-of-the-art techniques. While our heuristics produce shallow circuits compared to existing approaches, they still require transpilation to adhere to hardware coupling constraints. Typically, this involves inserting SWAP gates to route logical qubits across the physical qubit layout, resulting in increased entangling gate overhead. There is optimism that our approach, which generates shallower and more compact circuits, will ultimately reduce overall circuit depth.

An alternative, more promising approach involves directly adapting these heuristics to the constraints of the target architecture. This adaptation will be the focus of future research efforts.

\section*{Acknowledgements}

The authors want to thank Ali Javadi for proof reading the paper and Arya Maheshwari for wrapping the Paulihedral code. This work has been co-funded by the European Commission as part of the EIC accelerator program under the grant agreement 190188855 for SEPOQC project.

\bibliography{biblio}
\bibliographystyle{abbrv}

\end{document}

%% file: table_new_4.tex
\begin{table}[]
    \centering

\begin{tabular}{lrrrrr}
\toprule
 & naive & ph & tket & rdepth & rcount \\
 & count & count & count & count & count \\
name &  &  &  &  &  \\
\midrule
LiH cmplt JW sto3g & 38 & -36.8\% & -52.6\% & {\cellcolor{green}} -68.4\% & {\cellcolor{green}} -68.4\% \\
LiH frz P sto3g & 38 & -36.8\% & -57.8\% & {\cellcolor{green}} -76.3\% & -73.6\% \\
H2O cmplt P sto3g & 630 & -64.4\% & -67.6\% & -83.4\% & {\cellcolor{green}} -86.0\% \\
C2H4 cmplt JW sto3g & 666 & -59.9\% & -62.0\% & {\cellcolor{green}} -80.4\% & -80.0\% \\
H2 cmplt P ccpvdz & 768 & -72.5\% & -73.8\% & -86.7\% & {\cellcolor{green}} -87.8\% \\
LiH cmplt BK ccpvdz & 1448 & -70.5\% & -66.5\% & -82.5\% & {\cellcolor{green}} -85.2\% \\
H2 cmplt JW ccpvdz & 1316 & -56.2\% & -64.5\% & -78.5\% & {\cellcolor{green}} -85.1\% \\
H2O cmplt JW 631g & 1442 & -64.0\% & -69.6\% & -77.8\% & {\cellcolor{green}} -88.1\% \\
C2H4 frz JW sto3g & 1440 & -67.6\% & -71.4\% & -79.8\% & {\cellcolor{green}} -84.3\% \\
H2O cmplt P 631g & 1616 & -77.1\% & -76.3\% & -82.6\% & {\cellcolor{green}} -90.1\% \\
NH frz P sto3g & 1316 & -60.7\% & -65.9\% & -76.6\% & {\cellcolor{green}} -82.2\% \\
H4 cmplt JW ccpvdz & 3700 & -60.0\% & -64.8\% & -75.1\% & {\cellcolor{green}} -82.3\% \\
LiH cmplt BK 631g & 3896 & -69.1\% & -75.7\% & -78.1\% & {\cellcolor{green}} -85.0\% \\
H4 cmplt BK ccpvdz & 4178 & -67.4\% & -71.3\% & -80.2\% & {\cellcolor{green}} -84.6\% \\
H2 cmplt JW 631g & 7640 & -72.0\% & -72.1\% & -71.8\% & {\cellcolor{green}} -80.7\% \\
LiH frz JW ccpvdz & 8064 & -78.1\% & -78.6\% & -75.8\% & {\cellcolor{green}} -82.5\% \\
LiH cmplt BK sto3g & 7796 & -64.1\% & -70.2\% & -72.8\% & {\cellcolor{green}} -80.7\% \\
LiH frz BK sto3g & 8064 & -70.2\% & -78.6\% & -74.2\% & {\cellcolor{green}} -83.0\% \\
NH frz P 631g & 8004 & -69.0\% & -73.4\% & -75.2\% & {\cellcolor{green}} -81.1\% \\
H8 cmplt JW sto3g & 8004 & -69.0\% & -73.4\% & -75.2\% & {\cellcolor{green}} -81.1\% \\
H2 cmplt P ccpvtz & 8680 & -75.2\% & -75.8\% & -79.2\% & {\cellcolor{green}} -83.0\% \\
C2H4 frz BK sto3g & 11344 & -78.5\% & -77.6\% & -82.7\% & {\cellcolor{green}} -92.8\% \\
H4 cmplt BK sto3g & 11344 & -78.5\% & -77.6\% & -82.7\% & {\cellcolor{green}} -92.8\% \\
H2 cmplt BK ccpvdz & 10344 & -74.8\% & -76.2\% & -74.4\% & {\cellcolor{green}} -82.2\% \\
\bottomrule
\end{tabular}
    \caption{Entangling count comparison for UCCSD Ansätze synthesis for different molecules, encodings, and bases. The first column displays the naive entangling gate count obtained by individually synthesizing each rotation using two CNOT ladders. Other columns display the entangling gate count gain with respect to that value. The best result of each row is outlined in green.}
    \label{tab:tab_chem_1}
\end{table}
\begin{table}[]
    \centering

\begin{tabular}{lrrrrr}
\toprule
 & naive & ph & tket & rdepth & rcount \\
 & count & count & count & count & count \\
name &  &  &  &  &  \\
\midrule
H2 cmplt P sto3g & 9938 & -68.0\% & -67.5\% & -74.0\% & {\cellcolor{green}} -80.3\% \\
H2O cmplt JW sto3g & 13878 & -85.5\% & -76.4\% & -86.6\% & {\cellcolor{green}} -94.9\% \\
LiH cmplt P ccpvdz & 10228 & -70.8\% & -68.2\% & -74.9\% & {\cellcolor{green}} -82.6\% \\
LiH frz BK 631g & 14616 & -87.5\% & -81.6\% & -86.4\% & {\cellcolor{green}} -94.6\% \\
H2O frz BK sto3g & 14616 & -87.5\% & -81.6\% & -86.4\% & {\cellcolor{green}} -94.6\% \\
LiH frz JW sto3g & 13108 & -70.4\% & -73.6\% & -71.8\% & {\cellcolor{green}} -80.0\% \\
C2H4 cmplt P sto3g & 14360 & -71.0\% & -79.9\% & -72.7\% & {\cellcolor{green}} -80.3\% \\
LiH frz BK ccpvdz & 14020 & -67.9\% & -70.5\% & -72.1\% & {\cellcolor{green}} -82.4\% \\
LiH cmplt P 631g & 17924 & -75 & -74.3\% & -69.7\% & {\cellcolor{green}} -81.4\% \\
H4 cmplt BK 631g & 19574 & -69.5\% & -69.2\% & -71.3\% & {\cellcolor{green}} -79.5\% \\
H8 cmplt BK sto3g & 21072 & -74.8\% & -78.9\% & -72.7\% & {\cellcolor{green}} -80.6\% \\
NH frz JW 631g & 22316 & -78.7\% & -76.1\% & -78.2\% & {\cellcolor{green}} -80.4\% \\
CH2 cmplt JW sto3g & 23200 & {\cellcolor{green}} -83.2\% & -81.9\% & -69.6\% & -81.0\% \\
CH2 cmplt BK sto3g & 37272 & -74.9\% & -74.1\% & -62.2\% & {\cellcolor{green}} -78.0\% \\
C2H4 frz P sto3g & 42368 & -77.4\% & {\cellcolor{green}} -80.5\% & -68.7\% & -80.0\% \\
H4 cmplt P sto3g & 41432 & -73.5\% & -69.6\% & -68.7\% & {\cellcolor{green}} -79.1\% \\
CH2 frz JW sto3g & 49880 & -83.4\% & -79.5\% & -72.8\% & {\cellcolor{green}} -94.2\% \\
H2 cmplt BK ccpvtz & 86598 & -92.1\% & -81.1\% & -79.1\% & {\cellcolor{green}} -96.8\% \\
H4 cmplt P ccpvdz & 89080 & -92.7\% & -84.1\% & -83.9\% & {\cellcolor{green}} -97.0\% \\
H2 cmplt BK sto3g & 53450 & -77.6\% & -76.0\% & -76.2\% & {\cellcolor{green}} -91.3\% \\
H2O cmplt BK sto3g & 67270 & -83.9\% & -79.0\% & -79.4\% & {\cellcolor{green}} -91.2\% \\
H4 cmplt JW sto3g & 86650 & {\cellcolor{green}} -80.0\% & -75.0\% & -55.5\% & -77.2\% \\
H2O cmplt BK 631g & 88536 & {\cellcolor{green}} -82.6\% & -82.2\% & -63.0\% & -77.0\% \\
NH cmplt BK 631g & 75664 & {\cellcolor{green}} -77.4\% & -72.3\% & -51.4\% & -74.3\% \\
\bottomrule
\end{tabular}
\caption{See table \ref{tab:tab_chem_1} for details}
    \label{tab:tab_chem_2}
\end{table}

\begin{table}[]
    \centering

\begin{tabular}{lrrrrr}
\toprule
 & naive & ph & tket & rdepth & rcount \\
 & count & count & count & count & count \\
name &  &  &  &  &  \\
\midrule
H2O frz JW 631g & 168986 & -88.8\% & -81.6\% & -55.5\% & {\cellcolor{green}} -96.6\% \\
H2 cmplt BK 631g & 173264 & -82.1\% & {\cellcolor{green}} -82.7\% & -55.9\% & -76.7\% \\
NH cmplt JW 631g & 170666 & {\cellcolor{green}} -79.9\% & -74.6\% & -53.3\% & -76.2\% \\
LiH cmplt P sto3g & 149540 & {\cellcolor{green}} -78.9\% & -73.8\% & -47.8\% & -72.8\% \\
LiH frz P ccpvdz & 335178 & -95.0\% & -83.3\% & -74.4\% & {\cellcolor{green}} -98.2\% \\
CH2 frz BK sto3g & 341280 & -95.2\% & -85.3\% & -77.5\% & {\cellcolor{green}} -98.2\% \\
H4 cmplt JW 631g & 244992 & -83.0\% & {\cellcolor{green}} -85.7\% & -50.6\% & -75.3\% \\
LiH cmplt JW ccpvdz & 230810 & -81.6\% & -78.5\% & -58.8\% & {\cellcolor{green}} -94.0\% \\
C2H4 cmplt BK sto3g & 276768 & -83.8\% & -81.8\% & -46.8\% & {\cellcolor{green}} -93.6\% \\
H4 cmplt P 631g & 401046 & -90.0\% & -84.4\% & -77.4\% & {\cellcolor{green}} -96.6\% \\
H8 cmplt P sto3g & 407320 & -92.5\% & -87.0\% & -78.3\% & {\cellcolor{green}} -96.5\% \\
NH cmplt JW sto3g & 312096 & -81.1\% & {\cellcolor{green}} -82.1\% & -50.7\% & -76.6\% \\
NH frz BK sto3g & 309068 & {\cellcolor{green}} -78.8\% & -71.5\% & -56.1\% & -76.0\% \\
NH frz BK 631g & 260656 & {\cellcolor{green}} -79.5\% & -70.6\% & -47.5\% & -71.4\% \\
H2O frz P sto3g & 472132 & -90.4\% & -85.3\% & -74.7\% & {\cellcolor{green}} -96.3\% \\
NH frz JW sto3g & 479136 & -92.8\% & -88.2\% & -79.2\% & {\cellcolor{green}} -96.3\% \\
LiH frz JW 631g & 317504 & {\cellcolor{green}} -78.2\% & -67.4\% & -28.2\% & -67.8\% \\
NH cmplt P 631g & 410000 & {\cellcolor{green}} -81.3\% & -75.1\% & -45.6\% & -75.0\% \\
CH2 frz P sto3g & 414240 & -82.9\% & {\cellcolor{green}} -84.4\% & -44.3\% & -75.2\% \\
NH cmplt BK sto3g & 488548 & {\cellcolor{green}} -78.0\% & -73.7\% & -22.0\% & -66.9\% \\
H2 cmplt JW ccpvtz & 640768 & -82.6\% & {\cellcolor{green}} -86.3\% & -48.2\% & -74.3\% \\
H2 cmplt P 631g & 636668 & {\cellcolor{green}} -80.0\% & -75.5\% & -47.3\% & -74.5\% \\
\bottomrule
\end{tabular}
\caption{See table \ref{tab:tab_chem_1} for details}
    \label{tab:tab_chem_3}
\end{table}

\begin{table}[]
    \centering

\begin{tabular}{lrrrrr}
\toprule
 & naive & ph & tket & rdepth & rcount \\
 & depth & depth & depth & depth & depth \\
name &  &  &  &  &  \\
\midrule
LiH cmplt JW sto3g & 32 & -28.1\% & -46.8\% & {\cellcolor{green}} -68.7\% & -65.6\% \\
LiH frz P sto3g & 33 & -27.2\% & -57.5\% & {\cellcolor{green}} -81.8\% & -69.6\% \\
H2O cmplt P sto3g & 406 & -47.5\% & -53.9\% & {\cellcolor{green}} -88.6\% & -83.2\% \\
C2H4 cmplt JW sto3g & 409 & -36.4\% & -42.7\% & {\cellcolor{green}} -85.0\% & -75.7\% \\
H2 cmplt P ccpvdz & 436 & -53.4\% & -58.0\% & {\cellcolor{green}} -88.9\% & -84.8\% \\
LiH cmplt BK ccpvdz & 783 & -46.8\% & -42.1\% & {\cellcolor{green}} -86.9\% & -81.7\% \\
H2 cmplt JW ccpvdz & 796 & -32.1\% & -49.7\% & {\cellcolor{green}} -85.6\% & -81.2\% \\
H2O cmplt JW 631g & 810 & -37.7\% & -48.0\% & {\cellcolor{green}} -86.7\% & -85.5\% \\
C2H4 frz JW sto3g & 812 & -46.5\% & -60.4\% & {\cellcolor{green}} -86.4\% & -79.3\% \\
H2O cmplt P 631g & 830 & -56.7\% & -57.3\% & {\cellcolor{green}} -87.8\% & {\cellcolor{green}} -87.8\% \\
NH frz P sto3g & 838 & -42.1\% & -53.4\% & {\cellcolor{green}} -85.7\% & -78.8\% \\
H4 cmplt JW ccpvdz & 1944 & -30.9\% & -43.3\% & {\cellcolor{green}} -86.0\% & -77.2\% \\
LiH cmplt BK 631g & 1952 & -42.8\% & -60.6\% & {\cellcolor{green}} -86.5\% & -79.2\% \\
H4 cmplt BK ccpvdz & 2193 & -41.7\% & -50.4\% & {\cellcolor{green}} -88.4\% & -79.6\% \\
H2 cmplt JW 631g & 3727 & -47.6\% & -50.8\% & {\cellcolor{green}} -86.7\% & -73.2\% \\
LiH frz JW ccpvdz & 3764 & -56.9\% & -62.8\% & {\cellcolor{green}} -87.2\% & -74.6\% \\
LiH cmplt BK sto3g & 3766 & -34.6\% & -46.5\% & {\cellcolor{green}} -86.9\% & -72.3\% \\
LiH frz BK sto3g & 3768 & -40.6\% & -62.8\% & {\cellcolor{green}} -87.1\% & -75.9\% \\
NH frz P 631g & 3857 & -38.6\% & -51.0\% & {\cellcolor{green}} -87.5\% & -72.8\% \\
H8 cmplt JW sto3g & 3857 & -38.6\% & -51.0\% & {\cellcolor{green}} -87.5\% & -72.8\% \\
H2 cmplt P ccpvtz & 4225 & -52.0\% & -56.5\% & {\cellcolor{green}} -89.4\% & -75.8\% \\
C2H4 frz BK sto3g & 4867 & -50.5\% & -49.4\% & {\cellcolor{green}} -93.7\% & -89.2\% \\
H4 cmplt BK sto3g & 4867 & -50.5\% & -49.4\% & {\cellcolor{green}} -93.7\% & -89.2\% \\
H2 cmplt BK ccpvdz & 4908 & -52.1\% & -61.1\% & {\cellcolor{green}} -87.5\% & -75.5\% \\
\bottomrule
\end{tabular}
    
    \caption{Entangling depth comparison for UCCSD Ansätze synthesis for different molecules, encodings, and bases. The first column displays the naive entangling gate depth obtained by individually synthesizing each rotation using two CNOT ladders. Other columns display the entangling gate depth gain with respect to that value. The best result of each row is outlined in green.}
    \label{tab:tab_chem_depth_1}
\end{table}

\begin{table}[]
    \centering

\begin{tabular}{lrrrrr}
\toprule
 & naive & ph & tket & rdepth & rcount \\
 & depth & depth & depth & depth & depth \\
name &  &  &  &  &  \\
\midrule
H2 cmplt P sto3g & 4915 & -42.0\% & -46.2\% & {\cellcolor{green}} -87.7\% & -73.7\% \\
H2O cmplt JW sto3g & 5060 & -60.8\% & -37.9\% & {\cellcolor{green}} -94.1\% & -91.8\% \\
LiH cmplt P ccpvdz & 5160 & -46.3\% & -48.1\% & {\cellcolor{green}} -88.3\% & -76.6\% \\
LiH frz BK 631g & 5183 & -65.2\% & -50.3\% & {\cellcolor{green}} -93.9\% & -90.7\% \\
H2O frz BK sto3g & 5183 & -65.2\% & -50.3\% & {\cellcolor{green}} -93.9\% & -90.7\% \\
LiH frz JW sto3g & 6218 & -45.9\% & -52.5\% & {\cellcolor{green}} -88.4\% & -72.3\% \\
C2H4 cmplt P sto3g & 6240 & -37.6\% & -61.6\% & {\cellcolor{green}} -87.7\% & -70.0\% \\
LiH frz BK ccpvdz & 6246 & -37.4\% & -43.7\% & {\cellcolor{green}} -87.8\% & -74.0\% \\
LiH cmplt P 631g & 8412 & -52.5\% & -52.7\% & {\cellcolor{green}} -88.8\% & -74.4\% \\
H4 cmplt BK 631g & 9192 & -41.2\% & -51.4\% & {\cellcolor{green}} -88.3\% & -71.0\% \\
H8 cmplt BK sto3g & 9198 & -47.9\% & -61.6\% & {\cellcolor{green}} -88.2\% & -71.2\% \\
NH frz JW 631g & 9249 & -53.6\% & -50.4\% & {\cellcolor{green}} -90.6\% & -69.4\% \\
CH2 cmplt JW sto3g & 9295 & -61.7\% & -62.6\% & {\cellcolor{green}} -87.8\% & -69.7\% \\
CH2 cmplt BK sto3g & 16774 & -51.6\% & -55.0\% & {\cellcolor{green}} -87.3\% & -68.7\% \\
C2H4 frz P sto3g & 16984 & -50.1\% & -66.6\% & {\cellcolor{green}} -87.9\% & -68.2\% \\
H4 cmplt P sto3g & 16994 & -44.2\% & -46.3\% & {\cellcolor{green}} -88.1\% & -67.1\% \\
CH2 frz JW sto3g & 18176 & -55.1\% & -45.1\% & {\cellcolor{green}} -94.5\% & -92.3\% \\
H2 cmplt BK ccpvtz & 21541 & -68.7\% & -26.1\% & {\cellcolor{green}} -93.9\% & -93.5\% \\
H4 cmplt P ccpvdz & 21795 & -70.6\% & -36.8\% & {\cellcolor{green}} -95.0\% & -93.9\% \\
H2 cmplt BK sto3g & 22383 & -49.4\% & -49.4\% & {\cellcolor{green}} -92.5\% & -88.3\% \\
H2O cmplt BK sto3g & 22809 & -57.1\% & -46.4\% & {\cellcolor{green}} -92.2\% & -84.4\% \\
H4 cmplt JW sto3g & 30935 & -51.2\% & -46.5\% & {\cellcolor{green}} -86.0\% & -61.3\% \\
H2O cmplt BK 631g & 30970 & -55.6\% & -59.3\% & {\cellcolor{green}} -87.5\% & -59.9\% \\
NH cmplt BK 631g & 31739 & -50.4\% & -50.5\% & {\cellcolor{green}} -86.9\% & -62.9\% \\
\bottomrule
\end{tabular}

\caption{See table \ref{tab:tab_chem_depth_1} for details}
    \label{tab:tab_chem_depth_2}
\end{table}

\begin{table}[]
    \centering

\begin{tabular}{lrrrrr}
\toprule
 & naive & ph & tket & rdepth & rcount \\
 & depth & depth & depth & depth & depth \\
name &  &  &  &  &  \\
\midrule
H2O frz JW 631g & 51318 & -63.4\% & -40.0\% & {\cellcolor{green}} -93.9\% & -93.2\% \\
H2 cmplt BK 631g & 57380 & -53.3\% & -60.8\% & {\cellcolor{green}} -86.4\% & -58.5\% \\
NH cmplt JW 631g & 57410 & -49.1\% & -46.3\% & {\cellcolor{green}} -86.0\% & -58.1\% \\
LiH cmplt P sto3g & 58159 & -52.0\% & -52.0\% & {\cellcolor{green}} -86.4\% & -58.1\% \\
LiH frz P ccpvdz & 61396 & -72.8\% & -10.6\% & -94.3\% & {\cellcolor{green}} -94.5\% \\
CH2 frz BK sto3g & 61830 & -73.9\% & -20.3\% & {\cellcolor{green}} -94.8\% & -94.3\% \\
H4 cmplt JW 631g & 77014 & -54.2\% & -67.5\% & {\cellcolor{green}} -85.7\% & -54.6\% \\
LiH cmplt JW ccpvdz & 83151 & -52.7\% & -48.4\% & {\cellcolor{green}} -92.2\% & -91.0\% \\
C2H4 cmplt BK sto3g & 94580 & -56.1\% & -53.3\% & {\cellcolor{green}} -90.7\% & -90.1\% \\
H4 cmplt P 631g & 95435 & -61.8\% & -42.9\% & {\cellcolor{green}} -93.4\% & -92.5\% \\
H8 cmplt P sto3g & 95585 & -71.0\% & -51.6\% & {\cellcolor{green}} -93.5\% & -91.2\% \\
NH cmplt JW sto3g & 99877 & -53.3\% & -62.3\% & {\cellcolor{green}} -86.0\% & -57.3\% \\
NH frz BK sto3g & 101189 & -46.1\% & -43.0\% & {\cellcolor{green}} -87.4\% & -57.2\% \\
NH frz BK 631g & 101374 & -54.6\% & -49.7\% & {\cellcolor{green}} -87.3\% & -57.1\% \\
H2O frz P sto3g & 108634 & -62.3\% & -45.2\% & {\cellcolor{green}} -93.0\% & -91.5\% \\
NH frz JW sto3g & 108764 & -71.5\% & -55.2\% & {\cellcolor{green}} -93.9\% & -91.6\% \\
LiH frz JW 631g & 124158 & -51.5\% & -47.3\% & {\cellcolor{green}} -84.9\% & -53.0\% \\
NH cmplt P 631g & 125613 & -48.8\% & -45.1\% & {\cellcolor{green}} -85.3\% & -53.8\% \\
CH2 frz P sto3g & 126020 & -55.1\% & -62.8\% & {\cellcolor{green}} -85.0\% & -53.7\% \\
NH cmplt BK sto3g & 176403 & -48.3\% & -52.9\% & {\cellcolor{green}} -83.7\% & -49.1\% \\
H2 cmplt JW ccpvtz & 183920 & -53.1\% & -67.6\% & {\cellcolor{green}} -86.0\% & -50.6\% \\
H2 cmplt P 631g & 184436 & -45.8\% & -44.4\% & {\cellcolor{green}} -85.9\% & -51.1\% \\
\bottomrule
\end{tabular}
\caption{See table \ref{tab:tab_chem_depth_1} for details}
    \label{tab:tab_chem_depth_3}
\end{table}

%% file: table_resynth.tex
\begin{table}[]
    \centering
  
  \begin{tabular}{l|rr|rr|rr|r}
\toprule
 & \multicolumn{2}{c|}{init} & \multicolumn{2}{c|}{rcount} & \multicolumn{2}{c|}{rdepth} & \cite{Nam_2018, Amy_2018} \\
 name & count & depth & count & depth & count & depth & count \\
\midrule
adder-8 & 409 & 139 & {\cellcolor{green}} 289 & 177 & 415 & {\cellcolor{yellow}} 121 & 331 \\
barenco-tof-10 & 192 & 162 & 224 & 163 & 177 & {\cellcolor{yellow}} 110 & {\cellcolor{green}} 144 \\
barenco-tof-3 & 24 & 22 & 18 & 17 & {\cellcolor{green}} 16 & {\cellcolor{yellow}} 16 & 18 \\
barenco-tof-4 & 48 & 42 & {\cellcolor{green}} 33 & 28 & {\cellcolor{green}} 33 & {\cellcolor{yellow}} 27 & 36 \\
barenco-tof-5 & 72 & 62 & 58 & 50 & 55 & {\cellcolor{yellow}} 39 & {\cellcolor{green}} 54 \\
csla-mux-3 & 80 & 38 & 78 & 43 & {\cellcolor{green}} 69 & {\cellcolor{yellow}} 26 & 76 \\
csum-mux-9 & 168 & 32 & {\cellcolor{green}} 139 & 60 & 147 & {\cellcolor{yellow}} 27 & 148 \\
gf2-10-mult & 609 & 172 & 623 & 361 & 953 & {\cellcolor{yellow}} 152 & {\cellcolor{green}} 609 \\
gf2-16-mult & 1581 & {\cellcolor{yellow}} 292 & 1659 & 1008 & 3131 & 493 & {\cellcolor{green}} 1581 \\
gf2-4-mult & 99 & 58 & {\cellcolor{green}} 97 & 66 & 114 & {\cellcolor{yellow}} 38 & 99 \\
gf2-5-mult & 154 & 77 & 156 & 100 & 175 & {\cellcolor{yellow}} 55 & {\cellcolor{green}} 154 \\
gf2-6-mult & 221 & 96 & 226 & 153 & 305 & {\cellcolor{yellow}} 73 & {\cellcolor{green}} 221 \\
gf2-7-mult & 300 & 115 & 318 & 213 & 378 & {\cellcolor{yellow}} 80 & {\cellcolor{green}} 300 \\
gf2-8-mult & 405 & {\cellcolor{yellow}} 140 & 418 & 258 & 572 & 144 & {\cellcolor{green}} 405 \\
gf2-9-mult & 494 & 153 & 530 & 301 & 755 & {\cellcolor{yellow}} 123 & {\cellcolor{green}} 494 \\
grover-5 & 288 & 248 & 272 & {\cellcolor{yellow}} 226 & 414 & 232 & {\cellcolor{green}} 226 \\
ham15-high & 2149 & 1633 & 2303 & 1731 & 3185 & {\cellcolor{yellow}} 1359 & {\cellcolor{green}} 1502 \\
ham15-low & 236 & 167 & {\cellcolor{green}} 199 & 138 & 230 & {\cellcolor{yellow}} 120 & 208 \\
ham15-med & 534 & 418 & 362 & 279 & 549 & {\cellcolor{yellow}} 255 & {\cellcolor{green}} 357 \\
hwb6 & 116 & 94 & 109 & 92 & {\cellcolor{green}} 108 & {\cellcolor{yellow}} 65 & 110 \\
hwb8 & 7129 & 5014 & 8083 & 6081 & 10593 & {\cellcolor{yellow}} 4341 & {\cellcolor{green}} 6861 \\
mod-adder-1024 & 1720 & {\cellcolor{yellow}} 1336 & 3118 & 2145 & 4640 & 1476 & {\cellcolor{green}} 1390 \\
mod-mult-55 & 48 & 28 & 39 & 33 & {\cellcolor{green}} 36 & {\cellcolor{yellow}} 20 & 40 \\
mod-red-21 & 105 & 82 & 109 & 83 & 108 & {\cellcolor{yellow}} 67 & {\cellcolor{green}} 81 \\
mod5-4 & 28 & 28 & {\cellcolor{green}} 11 & {\cellcolor{yellow}} 9 & 13 &  10 & 26 \\
qcla-adder-10 & 233 & {\cellcolor{yellow}} 45 & {\cellcolor{green}} 186 & 68 & 208 & {\cellcolor{yellow}} 45 & 195 \\
qcla-com-7 & 186 & 49 & {\cellcolor{green}} 123 & 59 & 165 & {\cellcolor{yellow}} 48 & 132 \\
qcla-mod-7 & 382 & {\cellcolor{yellow}} 120 & 380 & 223 & 836 & 157 & {\cellcolor{green}} 302 \\
qft-4 & 46 & 43 & {\cellcolor{green}} 34 & {\cellcolor{yellow}} 32 & 50 & 39 & 48 \\
rc-adder-6 & 93 & {\cellcolor{yellow}} 55 & 72 & 58 & 85 & {\cellcolor{yellow}} 55 & {\cellcolor{green}} 71 \\
tof-10 & 102 & 86 & 85 & 63 & 81 & {\cellcolor{yellow}} 59 & {\cellcolor{green}} 70 \\
tof-3 & 18 & 16 & 14 & {\cellcolor{yellow}} 12 & {\cellcolor{green}} 13 & {\cellcolor{yellow}} 12 & 14 \\
tof-4 & 30 & 26 & {\cellcolor{green}} 22 & {\cellcolor{yellow}} 18 & {\cellcolor{green}} 22 & {\cellcolor{yellow}} 18 & {\cellcolor{green}} 22 \\
tof-5 & 42 & 36 & 36 & 32 & 33 & {\cellcolor{yellow}} 26 & {\cellcolor{green}} 30 \\
vbe-adder-3 & 70 & 49 & {\cellcolor{green}} 41 & 31 & 44 & {\cellcolor{yellow}} 27 & 46 \\
\bottomrule
\end{tabular}

    \caption{Compilation of standard circuits. The {\bf init} columns refer to the initial entangling count and depth of the circuits, computed by replacing every Toffoli gate by a standard 6 CNOTs template described in \cite{Nam_2018}. Last column contains the best CNOT count related in Table 1 of \cite{Amy_2018}: this number was achieved either by \cite{Amy_2018} or \cite{Nam_2018}. Both \cite{Amy_2018} and \cite{Nam_2018} only report CNOT count. The best count for each circuit is outlined in green. The best depth for each circuit is outlined in yellow.}
    \label{tab:lit_circuits}
\end{table}